\documentclass[runningheads]{llncs}
\usepackage[T1]{fontenc}
\usepackage{graphicx}
\usepackage{url}
\usepackage{color}

\usepackage{hyperref}
\hypersetup{
    colorlinks=true,
    linkcolor=blue,
    filecolor=blue,      
    urlcolor=blue,
    citecolor=blue,
}

\usepackage{caption} 
\usepackage{listings}
\usepackage{float}
\usepackage{appendix}

\usepackage{tabularx}
\usepackage{wrapfig}
\usepackage[ruled,vlined]{algorithm2e}
\usepackage{caption}
\usepackage{marvosym}

\usepackage{bbding}
\begin{document}

\title{Integrated Modeling, Verification, and Code Generation for Unmanned Aerial Systems}
%
%

\author{
Jianyu Zhang\inst{1} \and
Long Zhang\inst{2(}\textsuperscript{\normalsize{\Letter}}\inst{)} \and
Yixuan Wu\inst{3} \and 
Linru Ma\inst{2} \and
Feng Yang\inst{2} 
}
%

\institute{
School of Automation Engineering, University of Electronic Science and Technology of China, Chengdu, China \and
National Key Laboratory of Science and Technology on Information System Security, AMS, Beijing, China\\
\email{zhanglong10@nudt.edu.cn}
\and
School of Electronics and Information, Northwestern Polytechnical University, Xi’an, China \\
}
\maketitle             

\begin{abstract}
Unmanned Aerial Systems (UAS) are currently widely used in safety-critical fields such as industrial production, military operations, and disaster relief. Due to the diversity and complexity of application scenarios, UAS have become increasingly intricate. The challenge of designing and implementing highly reliable UAS while effectively controlling development costs and enhancing efficiency is a pressing issue faced by both academia and industry. Addressing this challenge, this paper aims to investigate an integrated approach to modeling, verification, and code generation for UAS. The paper begins by utilizing Architecture Analysis and Design Language (AADL) to model the UAS, proposing a set of generic UAS models. Based on these models, formal specifications are written to describe the system’s safety properties and functions. Finally, the paper introduces a method for generating flight controller code for UAS based on the verified models. Experiments conducted with the proposed method demonstrate its effectiveness in identifying potential vulnerabilities in the UAS during the early design phase and in generating viable flight controller code from the verified models. This approach can enhance the efficiency of designing and verifying high-reliability UAS.
\end{abstract}
\section{Introduction}\label{sec1}
UAS have become indispensable tools in many industries due to their enhanced processing capabilities, adaptability, and extended operational durations. \cite{gupta2013review,tan2020unmanned}. As the scenarios the systems need to adapt to become increasingly diverse, the complexity of UAS is on the rise \cite{balestrieri2021sensors}. Consequently, the safety and reliability of these systems are emerging as critical concerns.

UAS may encounter several security issues during operations, including hacking and component failures \cite{mohsan2023unmanned}. To address these issues, numerous security-enhancing technologies have been researched by both academia and industry. This paper categorizes these technologies based on their application stages into design-time security assurance technologies and runtime security assurance technologies.

Design-time security assurance technologies are primarily applied during the design phase of UAS and include in-the-loop simulation testing \cite{dai2021rflysim}, Failure Modes and Effects Analysis (FMEA) \cite{huang2020failure,shafiee2021unmanned}, and formal methods \cite{clarke1996formal}. The goal of these technologies is to identify and eliminate potential safety risks and ensure that system designs meet safety requirements. Runtime security assurance technologies are mainly used during the operation of UAS and include real-time monitoring \cite{witayangkurn2012real}, autonomous decision-making and control \cite{veres2011autonomous}, and Runtime Assurance (RTA) systems \cite{schierman2020runtime,lee1999runtime}. The aim of these technologies is to monitor the system’s state in real-time, respond promptly to emerging security issues, and ensure stability and safety during operation.

Currently, the limitation of design-time security assurance technologies is the high cost of verifying system reliability. These technologies need to deal with system complexity and uncertainty \cite{ferreira2010unmanned}, leading to a significant workload for verification. Runtime security assurance technologies also face challenges, such as balancing security and performance, managing complex system coordination and switching, and ensuring real-time system responses \cite{desai2019soter}.

Addressing these issues, this paper focuses on the design-time security assurance domain of UAS, aiming to enhance development and verification efficiency and reduce costs. The technical foundation of this research lies in architecture description languages \cite{medvidovic2000classification} and formal verification techniques \cite{clarke1996formal}. Architecture description languages are used to describe and represent system architectures, providing an abstract, high-level means to express system components, properties, and behaviors, supporting system design and analysis. AADL (Architecture Analysis and Design Language) \cite{feiler2006architecture} is one such commonly used architecture description language. Formal verification techniques aim to verify the correctness of software systems, hardware circuits, or protocols through rigorous mathematical methods, helping developers detect potential errors and security vulnerabilities.

This paper proposes an integrated method for modeling, verification, and code generation of UAS by combining architecture description languages and formal verification techniques. By synchronizing modeling and verification, the method assists developers in identifying potential system vulnerabilities during the system model design phase, enhancing system reliability and reducing verification costs. The code generation method then translates the verified models into flight controller code, improving development efficiency.

The main contributions of this paper are as follows:
\begin{itemize}
    \item Based on AADL, the paper develops a set of generic UAS models. These models include common components and properties of UAS and can facilitate the design and verification of unmanned systems.
    \item Using these models, the paper specifies and verifies common safety properties of UAS and designs verification algorithms for two safety functions. Verification can detect violations of specifications within the system and output specific counterexamples, aiding developers in identifying and remedying vulnerabilities.
    \item For the established models, the paper introduces a method for generating flight controller code for UAS, capable of directly producing viable flight controller code files from model files.
\end{itemize}

The structure of the remainder of this paper is as follows: Section \ref{sec2} introduces the related research. Section \ref{sec3} introduces the overall framework of our method. Section \ref{sec4} - \ref{sec6} introduce the three main parts of the method respectively. Section \ref{sec7} compares our method with other existing methods. Summary and future Outlook are provided in Section \ref{sec8}.

\section{Related Research}\label{sec2}
Unmanned Aerial Systems, as complete physical information systems, are often exposed to uncontrolled operational environments and face serious security threats. To ensure the safety and reliability of unmanned systems, extensive research has been conducted by both academia and industry.

Dai et al. \cite{dai2021rflysim} developed an automatic testing platform for UAS autopilot systems based on FPGA hardware-in-the-loop simulation and model-driven design, capable of simulating various failure modes and scenarios to verify and evaluate the accuracy and credibility of simulation models. Shafiee et al. \cite{shafiee2021unmanned} proposed a semi-quantitative reliability analysis framework based on FMEA to identify and assess the severity of failures in UAS during mission execution. Sadhu et al. \cite{sadhu2020board} introduced a deep learning-based method for detecting and identifying the causes of UAS failures by analyzing inertial measurement unit sensor data to detect abnormal behaviors and identify the causes of failures. Several studies have explored the application of formal verification techniques in unmanned systems, such as Luckcuck et al. \cite{luckcuck2023using}, who emphasized the role of formal methods as a key tool for ensuring the safety and correctness of unmanned systems, and believed that advancements in formal verification are crucial for addressing the challenges of verifying unmanned systems. Khan \cite{khan2020formal} used the formal proof tool Coq[23] to verify key hardware components of unmanned systems, highlighting the importance of formal verification of hardware components, especially in applications such as UAS and other computer-controlled systems.

The High-Assurance Cyber Military Systems (HACMS) project \cite{cofer2017secure} initiated by the Defense Advanced Research Projects Agency (DARPA) aims to research technologies for building highly reliable cyber-physical systems. The project developed a high-security UAS whose safety was proven using formal methods, and which effectively defended against malicious attacks from insiders. However, the project’s report \cite{fisher2017hacms} also mentioned its limitations, such as the significant workload and high costs in terms of time and manpower required for verification. The report also noted that verifying existing code is more difficult than simultaneous code development and verification, highlighting the need for research on methods of formal verification during the system development phase.

Among the aforementioned studies, methods based on in-the-loop simulation testing \cite{dai2021rflysim} and FMEA \cite{shafiee2021unmanned} struggle to provide comprehensive reliability proofs; the reliability of machine learning-based fault detection methods \cite{sadhu2020board,taimoor2023novel} in critical scenarios needs to be improved; and while formal method-based verification \cite{luckcuck2023using,khan2020formal,fisher2017hacms} can demonstrate system safety, the cost of verification is high. Addressing these issues, the objective of this paper is to propose a method that balances reliability with efficiency.

\section{Overall Framework of the Method}\label{sec3}
\begin{figure}
    \centering
    \includegraphics[width=\textwidth]{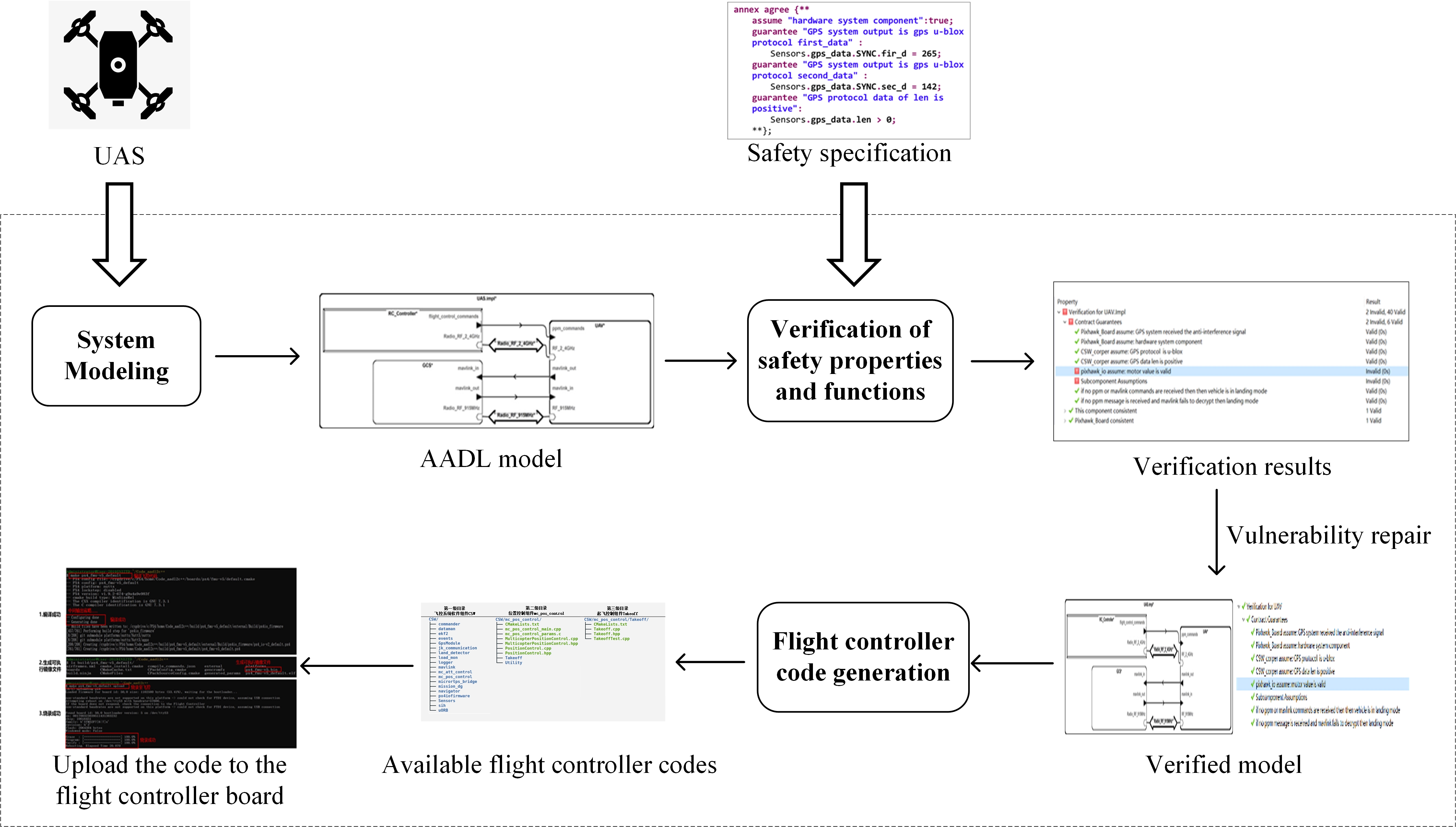}
    \caption{Overall framework of the method.}
    \label{fig1}
    \vspace{-10pt}
\end{figure}

This section introduces the overall framework of the method, as illustrated in Figure \ref{fig1}.

The integrated method for modeling, verification, and code generation of UAS consists of three main components: UAS modeling, verification of safety properties and functions, and flight controller code generation.

\begin{itemize}
    \item In the UAS modeling section, this paper establishes a set of system models for UAS based on AADL. The models, which are organized from top to bottom into multiple levels, can describe the overall structure and component properties of UAS. These models effectively support system design and verification. Additionally, the models are generic and can be applied to specific UAS models by modifying and refining certain components.
    \item In the verification of safety properties and functions section, this paper builds upon the established system models. Within the OSATE (Open Source AADL Tool Environment) environment \cite{feiler2006architecture}, the AGREE (Analysis for GNU Real-Time Extension) language \cite{whalen2012your} is used to describe common safety property specifications for UAS. For frequently used functions in UAS such as instruction encryption and GPS data security, the Resolute verification framework \cite{gacek2014resolute} is employed to design verification algorithms. After specifying the contracts, the paper conducts verification of the safety properties and functions, revealing potential vulnerabilities in the system.
    \item In the flight controller code generation section, this paper investigates the transformation rules from AADL code to flight controller code and designs the conversion method. This method enables the transformation of AADL system models into C++ code that can be used in the PX4 flight controller system \cite{meier2015px4}.    
\end{itemize}

The aforementioned work is implemented in the OSATE environment, an open-source integrated development environment that supports AADL. OSATE provides a complete toolkit for AADL, including an editor, analyzer, and code generator, to support the design, analysis, and implementation of AADL models \cite{feiler2004open}. Both AGREE and Resolute are formal verification tools integrated within OSATE. These tools offer assertion languages for verifying system safety and functionality, with Resolute being more suitable for scenarios that require detailed proofs.

\section{UAS Modeling}\label{sec4}
This section introduces the modeling method for Unmanned Aerial Systems, beginning with an analysis of the UAS architecture and then proceeding to model the system using AADL.

\subsection{UAS Architecture Analysis}
Unmanned Aerial Systems are integrated systems that comprise the drone, ground station, and the corresponding communication links. The drone is an aircraft without an onboard pilot, capable of autonomous flight or operation via remote control. The ground station is responsible for sending control commands to the drone and receiving data from it. UAS also require systems such as control and communication, as well as the necessary equipment and personnel to control the drone’s flight \cite{gupta2013review}.

To facilitate modeling, this paper categorizes the top-level components of the UAS into ground station component, remote controller component, and flight controller system component. The ground station component communicate with the drone’s flight controller system through a specific protocol, while the remote controller sends signals to the flight controller system via a receiver. These two components are primarily responsible for communication with the flight controller system. Therefore, in the modeling, we mainly describe their communication methods, connection relationships, and data interaction relationships without further refinement. The focus of our modeling refinement is on the flight controller system component.

The flight controller system component include hardware component, software component, and output control component. The hardware component describe the physical composition that ensures the normal operation of the drone’s flight control, including processors, sensors, power units, GPS devices, etc. The software component primarily describe the various process modules of the flight controller system, including software behaviors that respond to and process hardware devices, such as attitude control, position control, fusion algorithms, etc. The output control component mainly describe the process of control output, such as the conversion from PWM control signals to electronic speed controllers and motors.

\subsection{AADL Modeling Results}

Based on the description of the drone flight controller system provided in the previous section, the Unmanned Aerial System can be modeled using the AADL. AADL is a formal language designed to describe and validate the structure and behavior of embedded systems. Its design objective is to support multidimensional analysis of systems, including performance, safety, and reliability. AADL can appropriately ignore the specific implementations of components, modeling and validating embedded systems by describing the attributes of components and the interaction relationships between them. AADL can also model systems and components with a hierarchical structure at any necessary level of detail to assess various aspects of system performance \cite{feiler2006architecture}.

\begin{figure}
    \centering
    \includegraphics[width=0.6\textwidth]{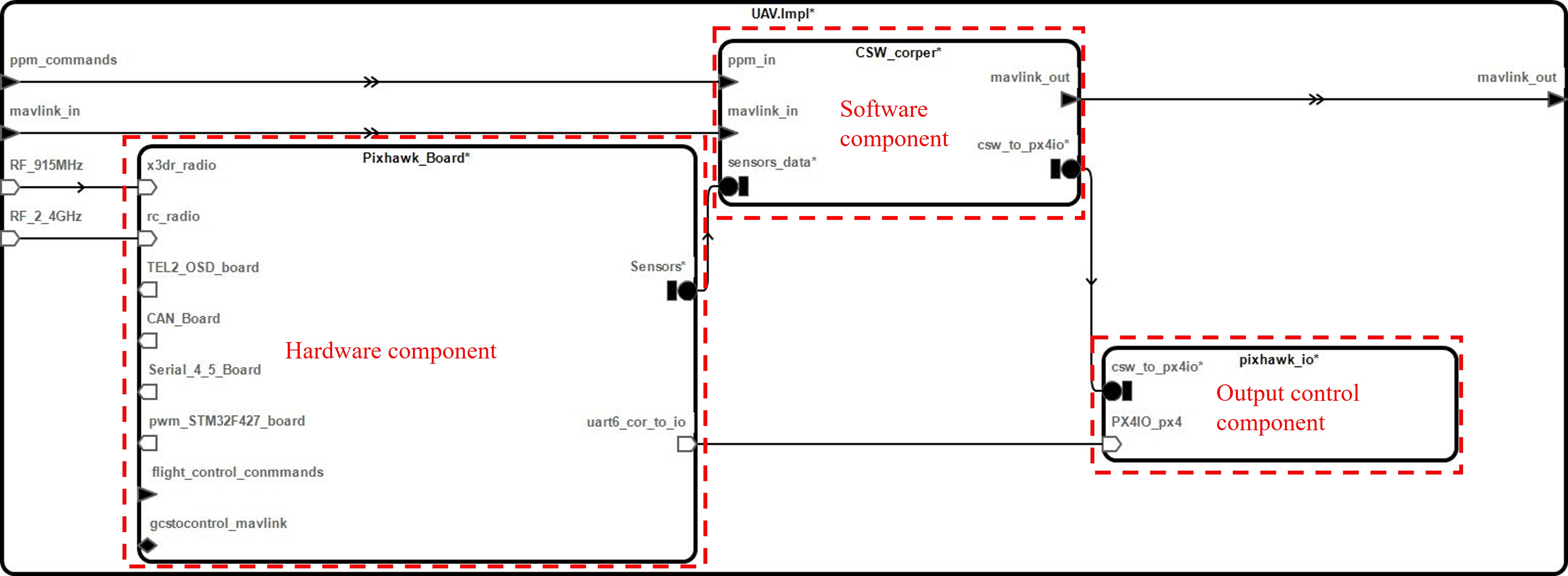}
    \caption{Top-level component structure.}
    \label{fig2}
    \vspace{-10pt}
\end{figure}

This paper models the system by describing the components at each layer of the drone system. In the model, the physical devices are abstracted into device components, and the data interfaces provided by each physical device are represented as ports in AADL. The software is placed within a system component, which contains multiple processes. The top-level component structure of the system model constructed in this paper is illustrated in Figure \ref{fig2}.

The top-level component consist of UAV, GCS, and RC\_Controller. UAV represents the flight controller system model of the drone, GCS is the ground control station system model, and RC\_Controller is the model of the drone’s remote controller. These three components communicate through specific signal lines.

\begin{figure}
    \centering
    \includegraphics[width=0.6\textwidth]{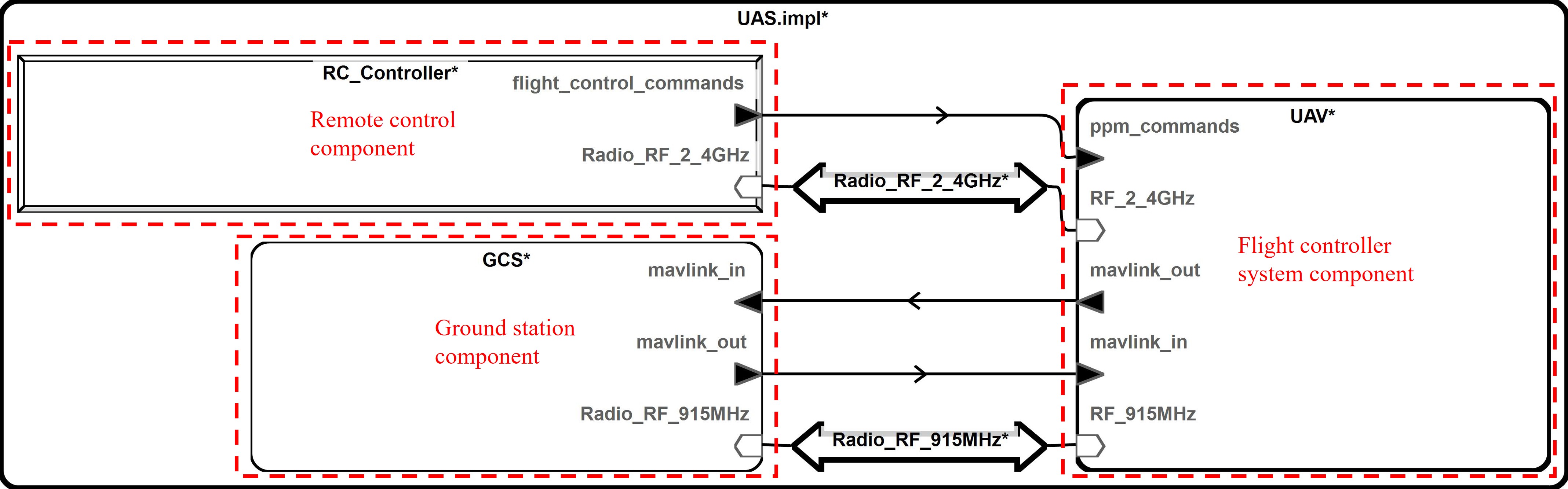}
    \caption{Flight controller system component structure.}
    \label{fig3}
    \vspace{-10pt}
\end{figure}

AADL supports multi-level modeling, and at the next level, the composition of the flight controller system component UAV is illustrated in Figure \ref{fig3}.

The flight controller system component further comprises sub-components such as the flight controller system hardware component (Pixhawk\_Board), software component (CSW\_corper), and I/O board component (pixhawk\_io). On the left side of the figure, the representation of different inputs to the supply components is shown. On the right side, the outputs are depicted.

\begin{figure}
    \centering
    \includegraphics[width=0.8\textwidth]{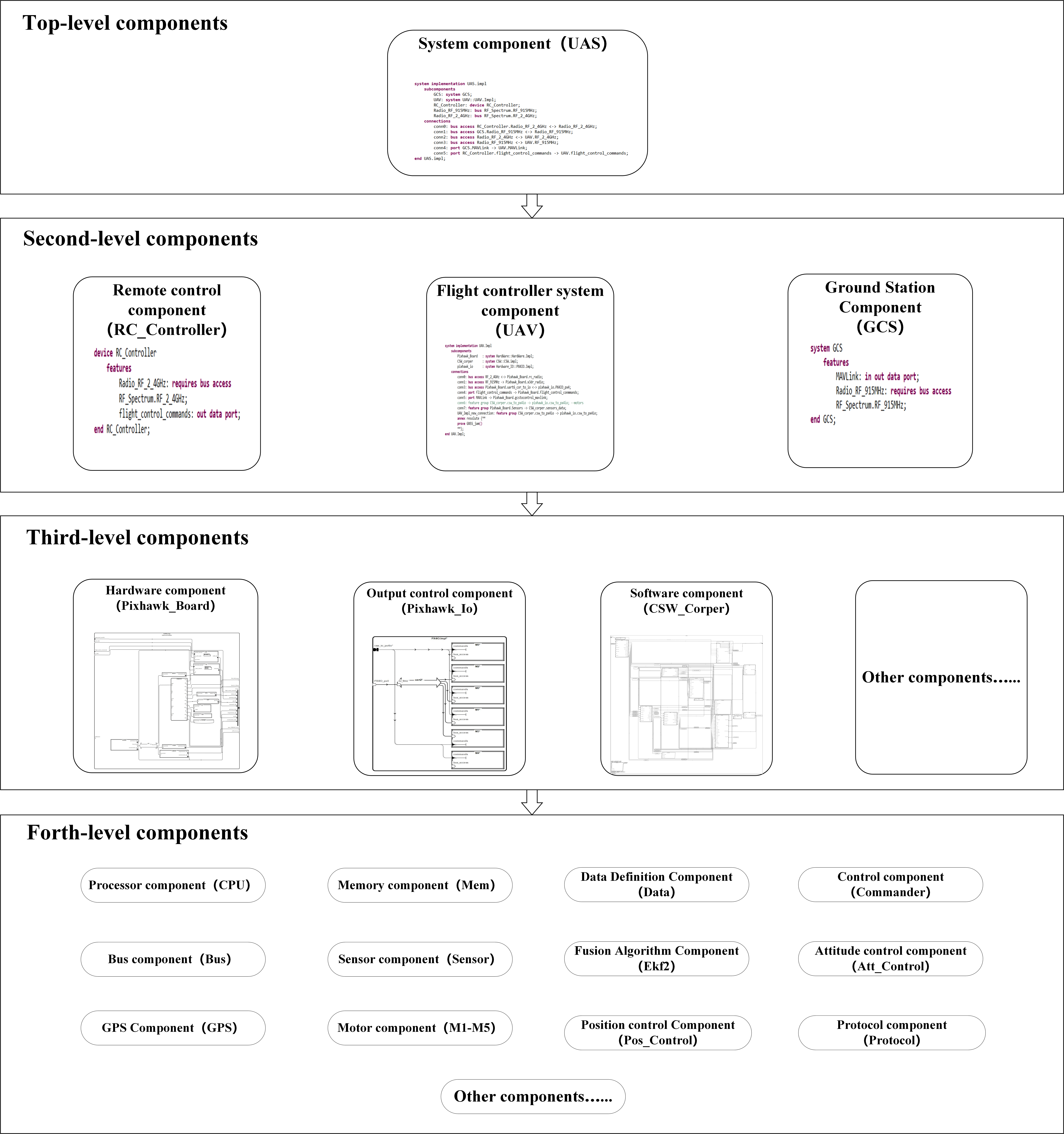}
    \caption{Model overall structure diagram.}
    \label{fig4}
    \vspace{-10pt}
\end{figure}

Through this layered modeling approach, this paper has completed the modeling of the Unmanned Aerial System, with the overall structure of the model illustrated in Figure \ref{fig4}.

The overall structure diagram shows the components at various levels of the system and the sub-components they contain. In the actual model, there are further refined components, but only up to the fourth level are listed here.

In each component, we have also defined its attributes, interfaces, and so on. Taking the sub-component ARM\_Cortex\_M7 in the processor component as an example.

\begin{figure}
    \centering
    \includegraphics[width=0.6\textwidth]{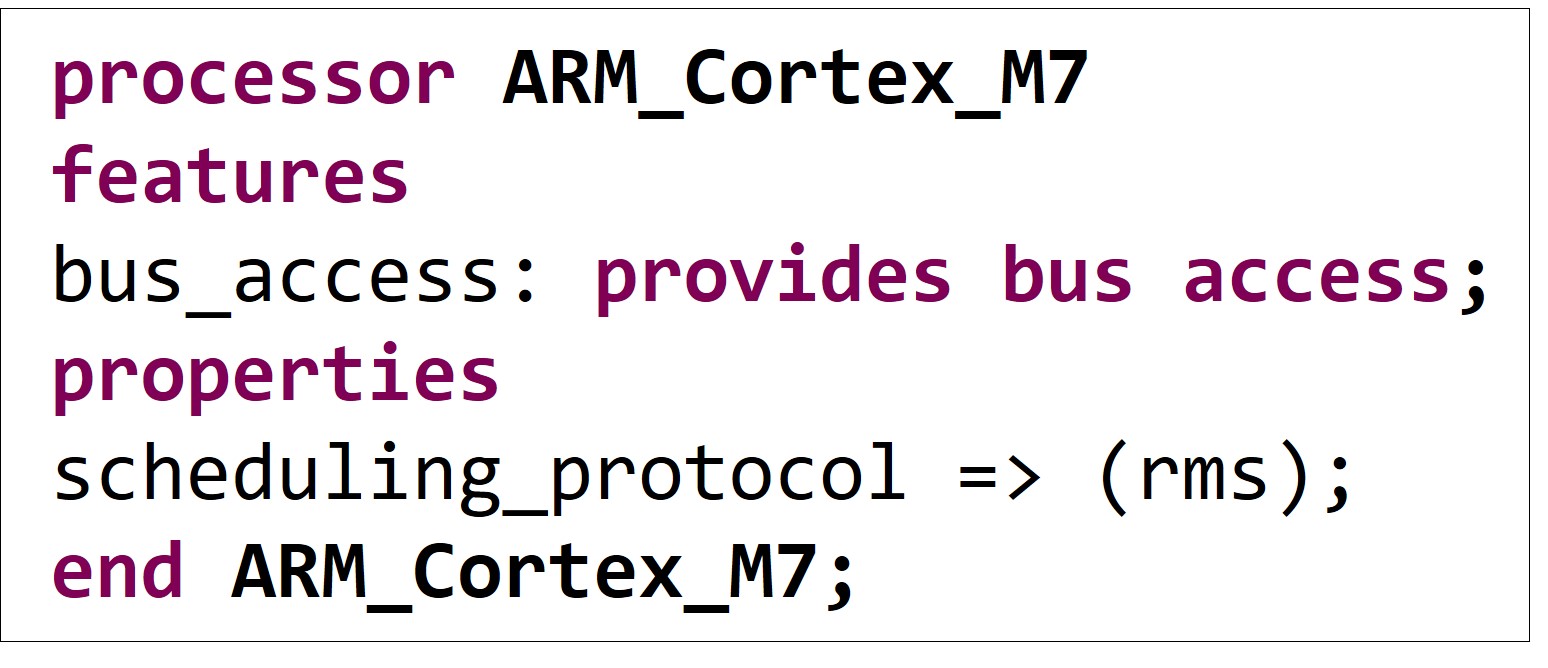}
    \caption{Subcomponent ARM\_Cortex\_M7 properties and interfaces.}
    \label{fig5}
    \vspace{-10pt}
\end{figure}

As illustrated in Figure \ref{fig5}, the above code describes a processor component named ARM\_Cortex\_M7, including its interface and attribute definitions. In AADL, the keyword processor is used to represent components that execute computational tasks. The features keyword defines the interfaces of the component, with bus access indicating that the processor provides bus access capabilities, enabling data transfer and communication. The properties keyword defines the attributes of the component, with scheduling\_protocol being a scheduling policy attribute, rms being a commonly used scheduling strategy in real-time systems. It determines priorities based on the task’s period, with shorter-period tasks having higher priority.

Using this model for design and verification offers the following advantages:
\begin{itemize}
    \item The model includes commonly used components and attributes of UAS, ensuring generality. Developers can modify or refine some components to adapt to specific configurations of UAS, thereby shortening the development cycle.
    \item Based on this model, verification of the components at various levels of the system can be conducted separately. In the event of vulnerabilities, modular repairs can be made; the overall functionality of the system can also be verified through the top-level components.   
\end{itemize}

\section{UAS Verification of Safety Properties and Functions}\label{sec5}
This section introduces the methods for verifying the safety properties and functions of UAS. It separately describes the specification of safety properties and the verification algorithms for security functions, followed by verification experiments and analysis.

\subsection{Specification of Safety Properties}
This paper employs the AGREE language to specify the safety properties of the UAS. AGREE is an analysis tool based on the AADL model, supporting the formal description and verification of system properties within AADL models. AGREE is well-suited for analyzing critical attributes of real-time systems, such as task scheduling, response time, and resource allocation. It helps ensure that the system operates correctly under specific constraints, thereby enhancing system reliability \cite{whalen2012your}.

\begin{figure}
    \centering
    \includegraphics[width=0.6\textwidth]{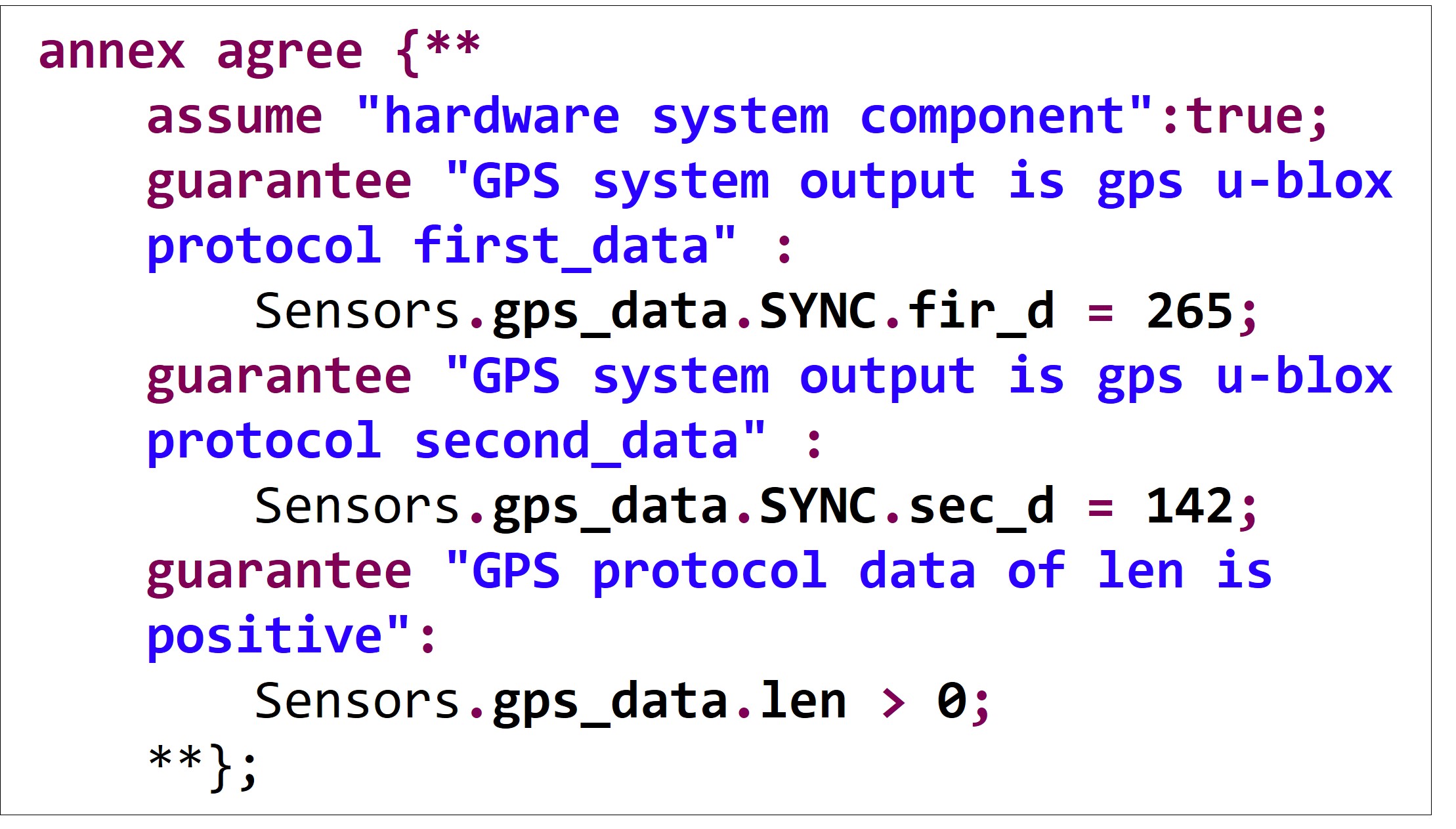}
    \caption{Security property specification example.}
    \label{fig6}
    \vspace{-10pt}
\end{figure}

Figure \ref{fig6} provides an example of safety property specification code for a flight controller system hardware component.

In the specification, three constraints are declared that the system must adhere to. The first and second fields of the GPS data structure are required to be 265 and 142, respectively, in accordance with the gps u-blox protocol. Additionally, the length field of the GPS data structure must be greater than 0. During system runtime verification, a search is conducted for the above data under various conditions. If counterexamples that do not meet the specifications are found, they are reported as vulnerabilities.

\begin{figure}
    \centering
    \includegraphics[width=0.75\textwidth]{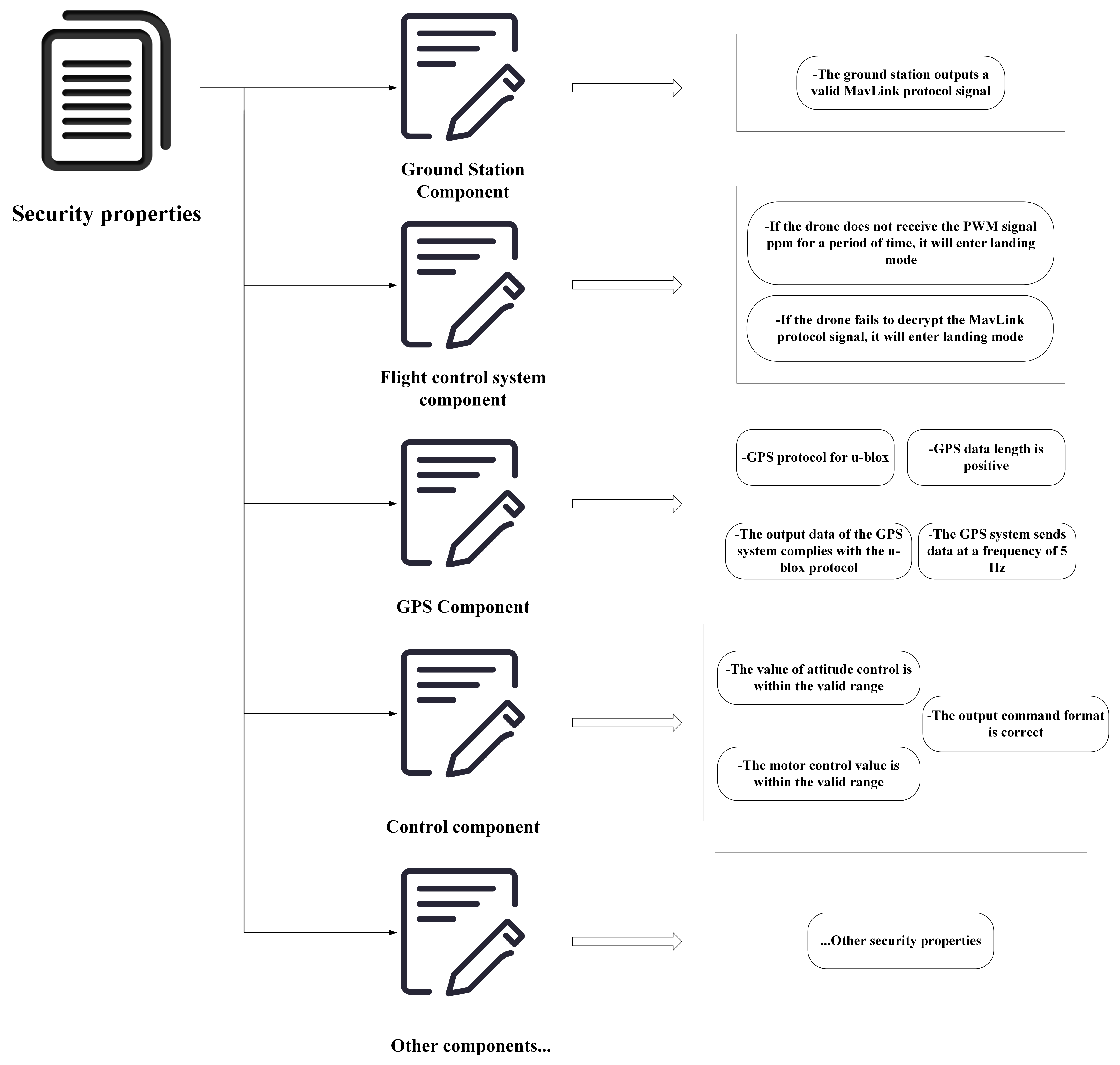}
    \caption{Description of safety properties of UAS.}
    \label{fig7}
    \vspace{-10pt}
\end{figure}

Similarly, this study describes common safety properties of the UAS, as illustrated in Figure \ref{fig7}.

This paper specifies safety properties at different levels of system components. Although it is challenging to consider all the necessary safety properties for a UAS during the design phase, this method allows for an incremental addition of required safety properties to each component, thereby enhancing system reliability.

\subsection{Verification Algorithms for Safety Functions}
This section verifies the instruction encryption and GPS data security functions of the UAS and designs corresponding verification algorithms.

The implementation of the verification algorithms uses the Resolute framework, which is a tool for verifying AADL models. Resolute provides an assertion language that allows developers to express system properties and constraints, and supports logical proofs of these properties. Resolute’s primary purpose is to rigorously evaluate the key properties of a system during the design phase, helping to ensure the reliability and safety of embedded systems \cite{gacek2014resolute}. Compared to AGREE, Resolute has stronger capabilities for logical proof, making it more suitable for scenarios that require detailed proofs and logical validation.

\noindent\textbf{Verification Algorithm for Instruction Encryption Function}.The verification algorithm for the instruction encryption function mainly describes a mechanism for secure instruction protection. It ensures that only encrypted instructions are accepted by the motor components, thereby preventing unauthorized access or tampering. Specifically, this mechanism includes the following steps:
\begin{itemize}
    \item Instruction Verification: First, check that the instruction is from the ground station and has been encrypted.
    \item Verification of Encryption Algorithm: Validate the effectiveness of the encryption algorithm.  
    \item Motors Only Accept Encrypted Instructions: Traverse each motor interface to ensure that the motors only accept and execute encrypted instructions.  
\end{itemize}

The specific description of the algorithm is shown in Algorithm \ref{Algorithm1}.

\begin{table}
    \centering
    \caption{Algorithm1: Verification Algorithm for GPS Data Security Function}
    \includegraphics[width=\textwidth]{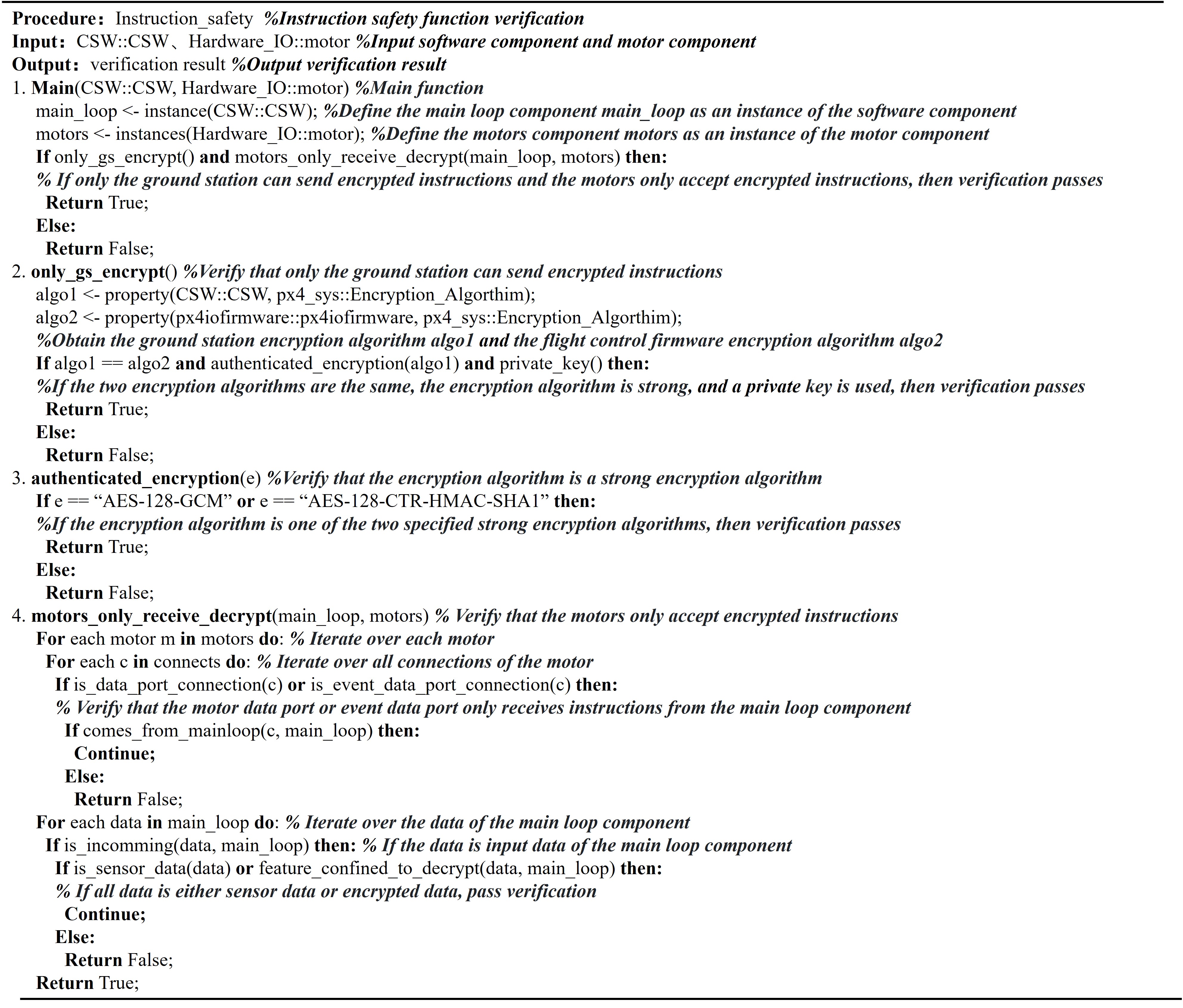}
    \label{Algorithm1}
    \vspace{10pt}
\end{table}

\noindent\textbf{Verification Algorithm for GPS Data Security Function}.The goal of this algorithm is to verify the security of GPS data, ensuring the integrity and security of the data during transmission and processing. The specific steps include:
\begin{itemize}
    \item Input/Output Validity Check: Verify if the trajectory history and output history are linear to ensure the continuity and consistency of the data.
    \item GPS Data Verification: Check each input data of the software component. Ensure that all GPS data reaching the software component is encrypted.
\end{itemize}

The specific description of the algorithm is shown in Algorithm \ref{Algorithm2}.

\begin{table}
    \centering
    \caption{Algorithm2:Verification Algorithm for GPS Data Security Function}\
    \includegraphics[width=\textwidth]{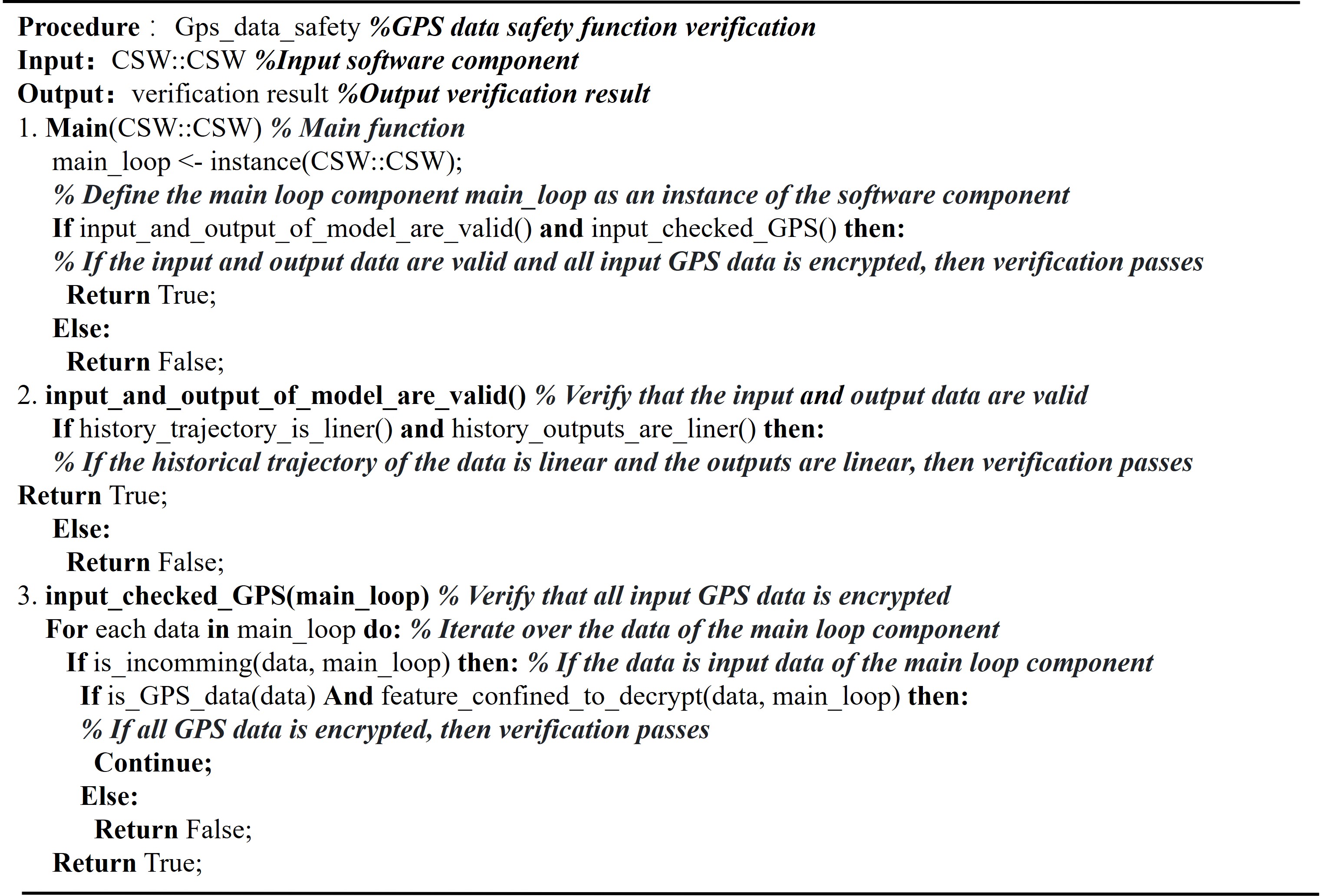}
    \label{Algorithm2}
    \vspace{-10pt}
\end{table}

\subsection{Verification Results and Analysis}
After completing the system modeling, safety property specifications, and functional verification algorithms, we executed the verification using the Z3 solver integrated within the OSATE environment. Through the verification process, several violations of the specifications were discovered, as shown in the table \ref{table1}.

\begin{table}
    \centering
    \caption{Vulnerability List.}\
    \includegraphics[width=\textwidth]{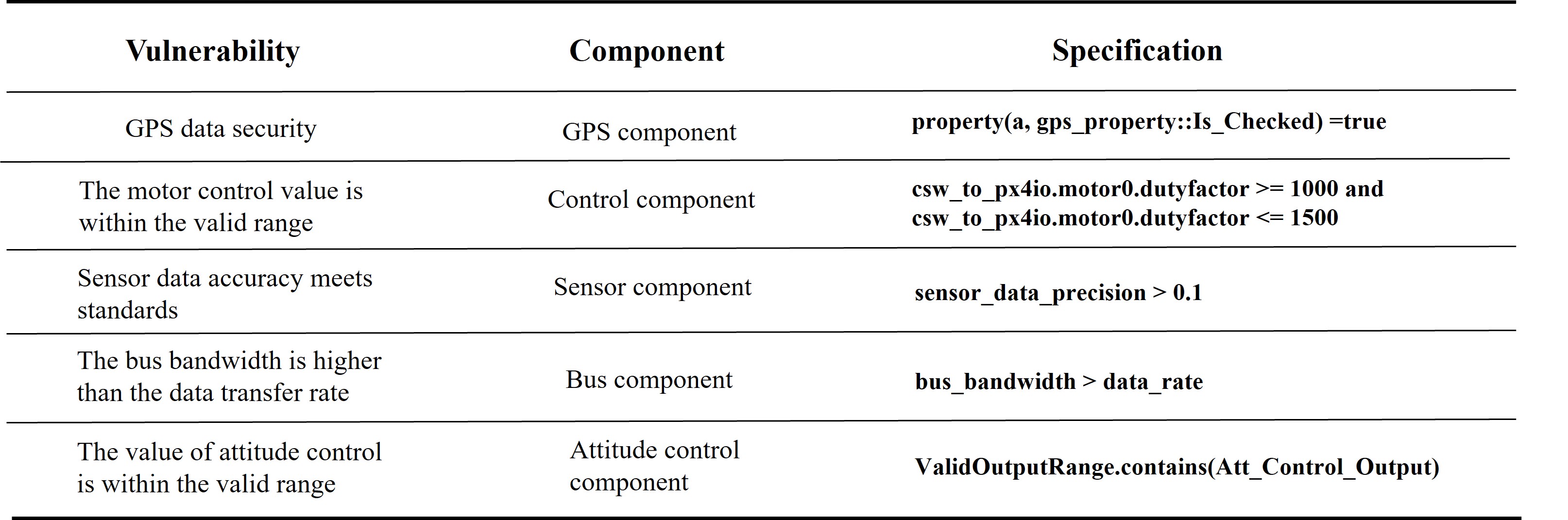}
    \label{table1}
    \vspace{-10pt}
\end{table}

\begin{figure}
    \centering
    \includegraphics[width=\textwidth]{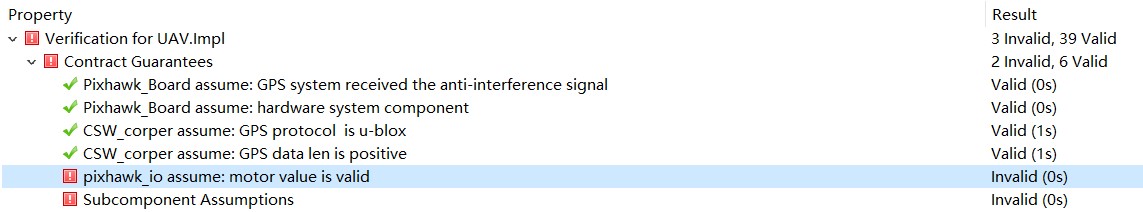}
    \caption{Validation results.}
    \label{fig8}
    \vspace{-10pt}
\end{figure}

After discovering vulnerabilities, the reasons for these vulnerabilities can be analyzed within the OSATE environment. Taking the specification “The value of motor control is within the effective range” as an example, the verification output results are illustrated in Figure \ref{fig8}.

\begin{figure}
    \centering
    \includegraphics[width=\textwidth]{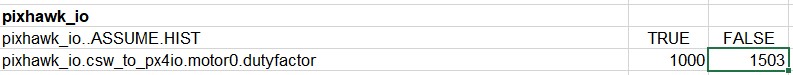}
    \caption{Counterexample Tracking.}
    \label{fig9}
    \vspace{-10pt}
\end{figure}

The OSATE environment also provides a feature for tracing counterexamples, as shown in Figure \ref{fig9}. For the specification violated in the above figure, it can be traced that the verification fails when the value of csw\_to\_px4io.motor0-5.dutyfactor (the pulse width modulation signal duty cycle for motor 0) is 1503.

In the specification, the range of this variable is limited to 1000-1500, which is clearly violated in this case. To resolve this vulnerability, we traced back to the location in the control component where the variable is assigned its value. It was found that within the control component, the range of assignment for this variable is 1000-2000, which is inconsistent with the specification. By correcting this range, the vulnerability can be fixed. Similar approaches can be used to resolve other vulnerabilities in the system.

Experiments have shown that modeling and verifying the UAS can help developers efficiently discover hidden vulnerabilities during the design phase, ensuring that the system meets the specified safety property specifications.

\section{UAS Flight Controller Code Generation}\label{sec6}
This section introduces the methods and experiments for generating flight controller code for UAS.

\subsection{Code Generation Method}
The code generation method involves a layer-by-layer transformation of the components. In the software component we have modeled, this includes system components, process components, thread components, data components, and subroutine components. This section primarily discusses the rules for converting each type of component.

\begin{figure}
    \centering
    \includegraphics[width=0.6\textwidth]{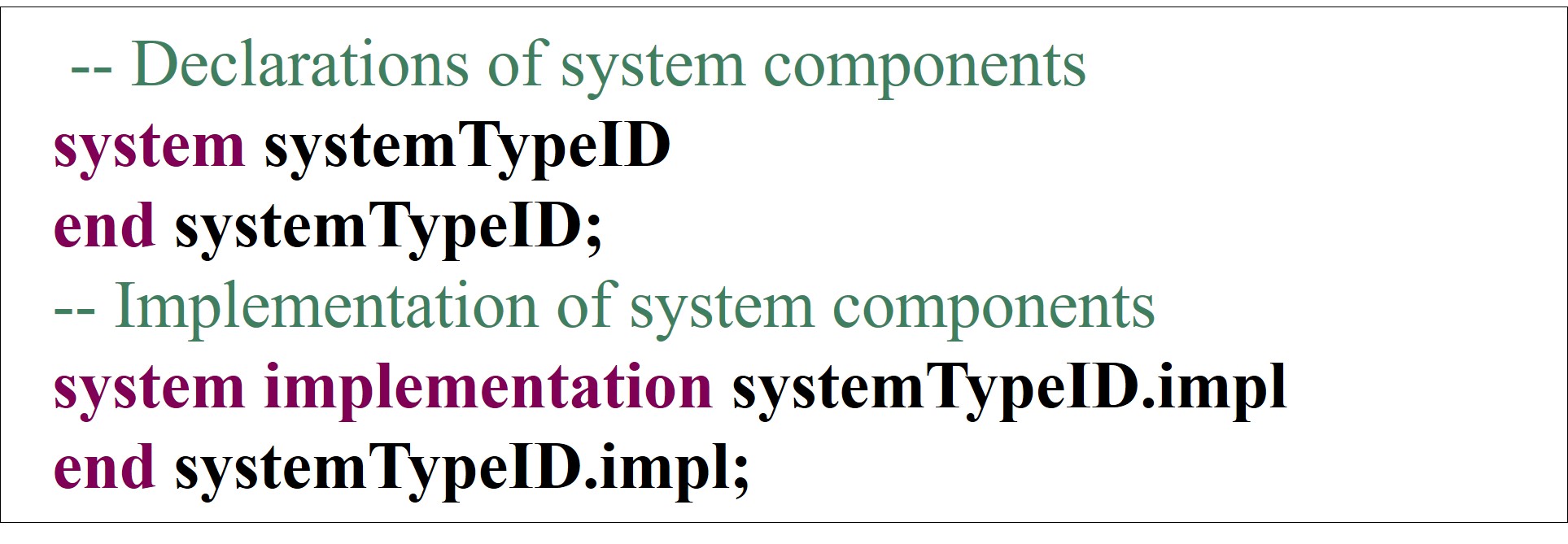}
    \caption{Declaration of system components.}
    \label{fig10}
    \vspace{-10pt}
\end{figure}

\noindent\textbf{System Component}.In AADL, the declaration of system components is illustrated in Figure \ref{fig10}, and system components should include both declaration and implementation.

The conversion rules for system components are as follows: System components are transformed into a folder named systemTypeID. This folder must contain at least one systemTypeID.hpp file, which includes declarations of global variables and functions, and includes common C and C++ language header files. systemTypeID is the identifier name for the system type declaration. All code files generated from the subcomponents of the system component are placed within the systemTypeID folder. This folder not only contains the code files generated from the system component but also includes all the code files generated from its subcomponents. For subcomponents, the system component will generate corresponding code files based on the specific conversion rules of the subcomponents.

\begin{figure}
    \centering
    \includegraphics[width=0.8\textwidth]{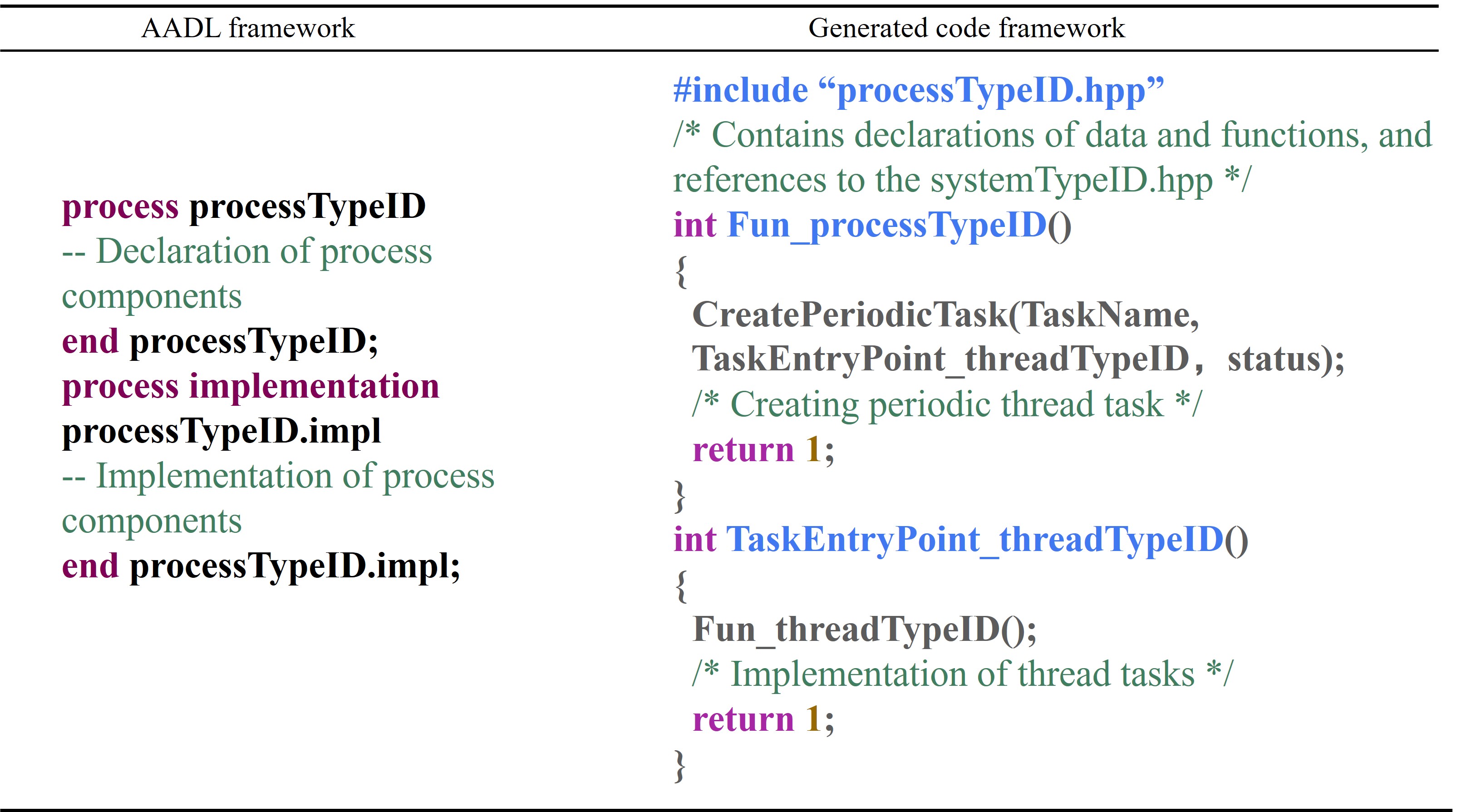}
    \caption{Process component conversion rules.}
    \label{fig11}
    \vspace{-10pt}
\end{figure}

\noindent\textbf{Process Component}.The conversion rules for process components are as follows: Each process subcomponent within the system component is transformed into a pair of processTypeID.hpp and processTypeID.cpp files. processTypeID.hpp contains declarations of data and functions shared by threads within the process and includes a reference to the systemTypeID.hpp file. processTypeID.cpp contains the implementations of the data and function declarations and includes the entry functions for the subcomponents of the process component: thread components and subroutine components. The conversion mapping is shown in Figure \ref{fig11}.

In the context of the code generation for UAS flight control systems, the function Fun\_processTypeID() serves as the entry point for process tasks. The function CreatePeriodicTask() is used to create a periodic task, with TaskName being the name of the task, TaskEntryPoint\_threadTypeID() being the entry function of the task, and status representing the status of the task. The function Fun\_threadTypeID() is the entry function for thread tasks within the process, indicating that once the process task is initialized, the thread subcomponent task is executed.

\begin{figure}
    \centering
    \includegraphics[width=0.8\textwidth]{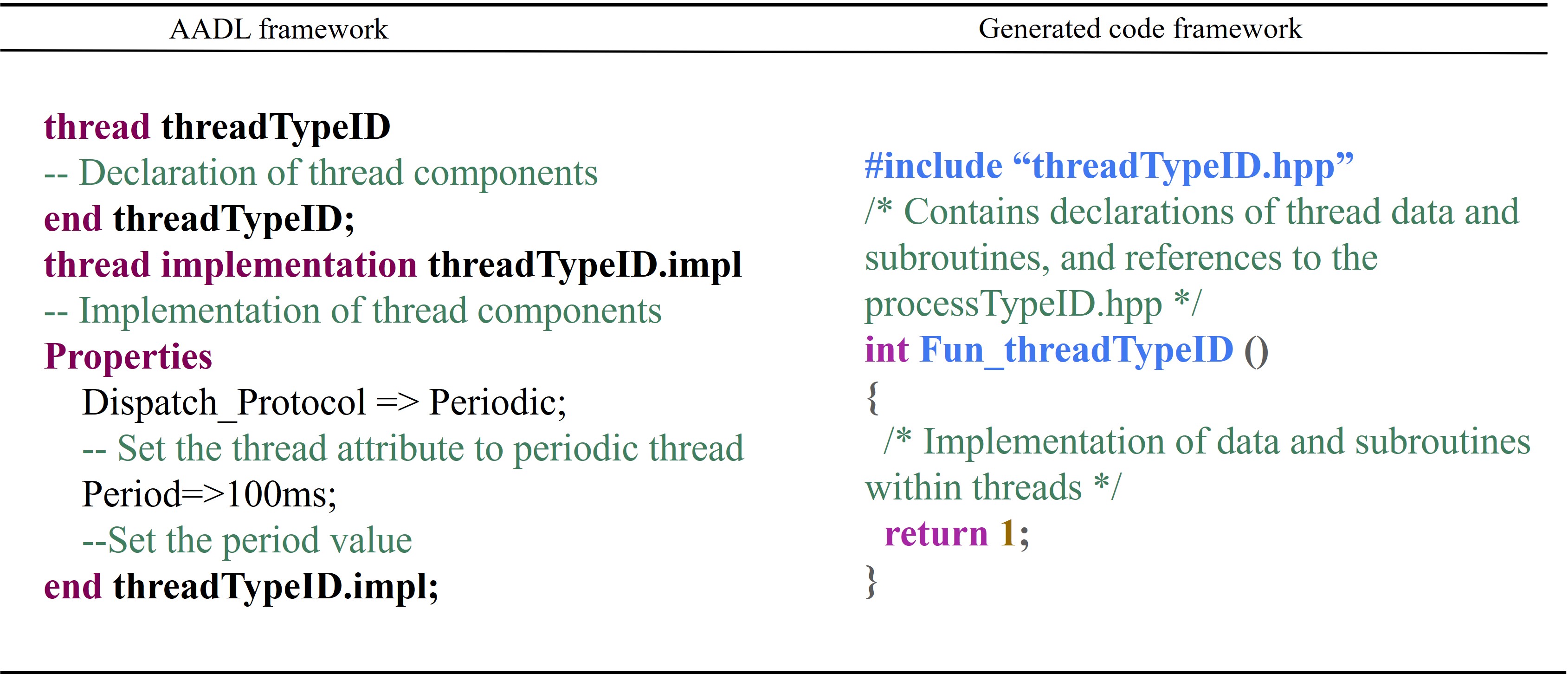}
    \caption{Thread component conversion rules.}
    \label{fig12}
    \vspace{-10pt}
\end{figure}

\noindent\textbf{Thread Component}.The conversion rules for thread components are as follows: Each thread subcomponent of the process component corresponds to a pair of threadTypeID.hpp and threadTypeID.cpp files. threadTypeID.hpp contains declarations of data and subroutines within the thread and includes the processTypeID.hpp file. threadTypeID.cpp contains the implementations of the data and subroutine declarations. The conversion mapping is shown in Figure \ref{fig12}.

\begin{figure}
    \centering
    \includegraphics[width=0.8\textwidth]{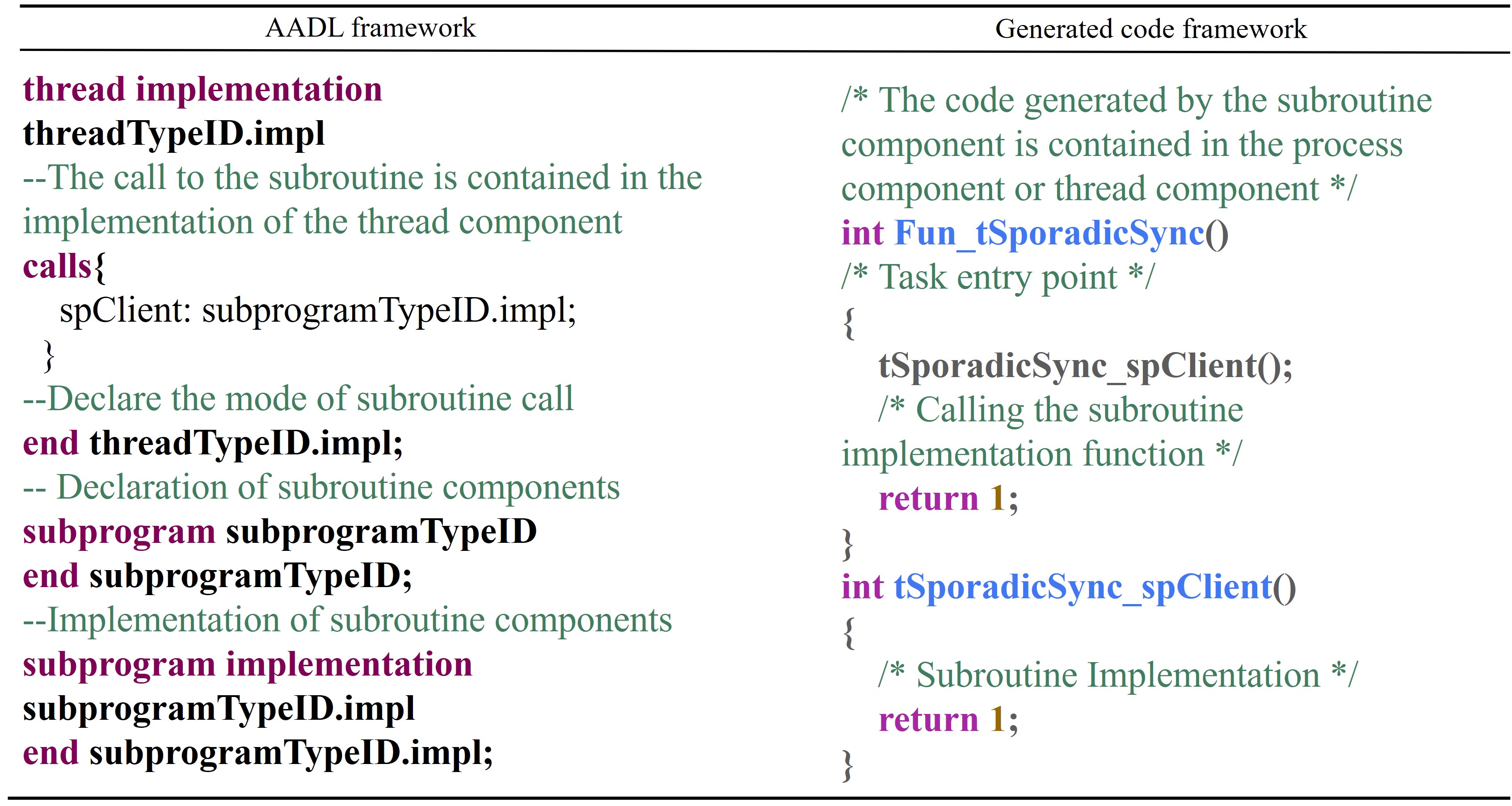}
    \caption{Subroutine component conversion rules.}
    \label{fig13}
    \vspace{-10pt}
\end{figure}

\noindent\textbf{Subroutine Component}.The conversion rules for subroutine components are as follows: Subroutine components are transformed into callable functions, and the generated code is included within the process component or thread component in which the subroutine resides. The conversion mapping is shown in Figure \ref{fig13}.

In this context, Fun\_tSporadicSync() is the entry point of the task, and tSporadicSync\_spClient() is the corresponding function for the subroutine, which can also include parameters.

\begin{figure}
    \centering
    \includegraphics[width=0.8\textwidth]{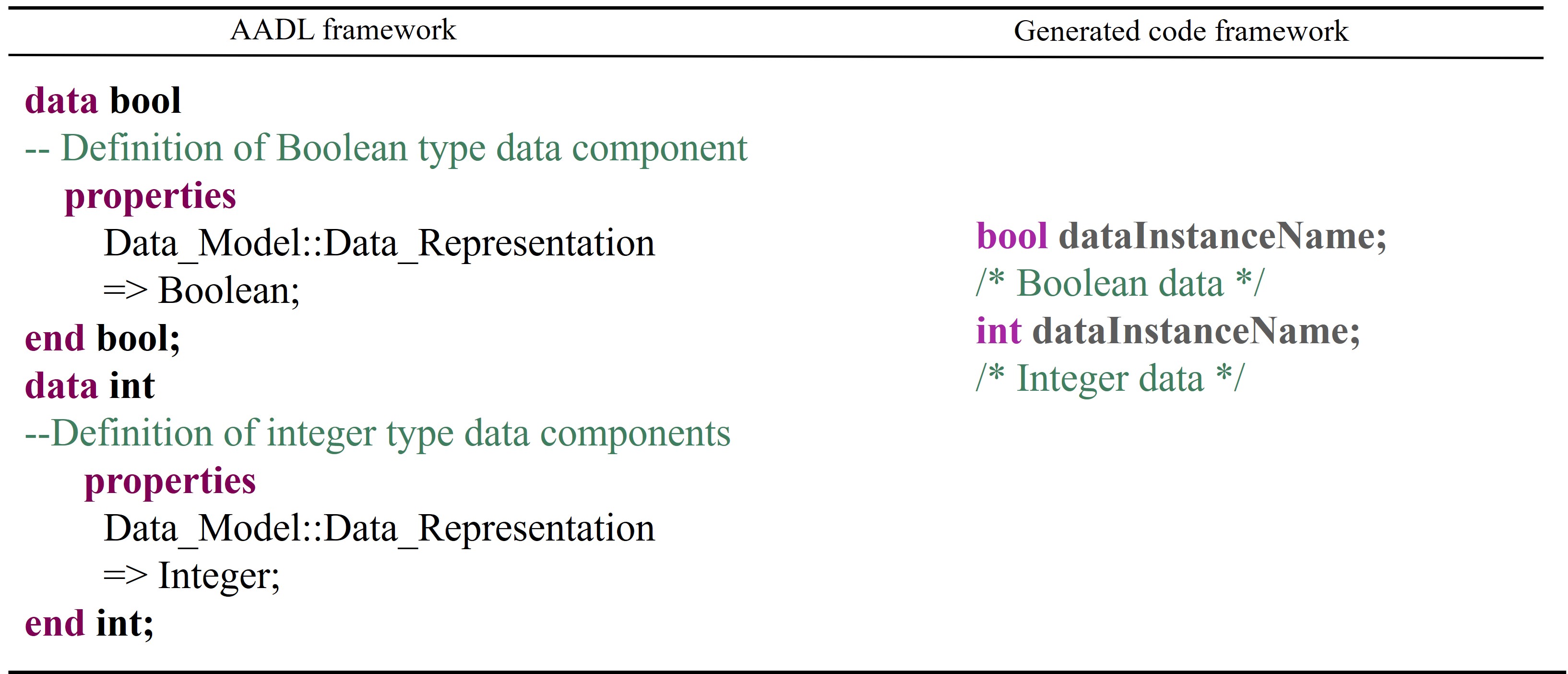}
    \caption{Data component conversion rules.}
    \label{fig14}
    \vspace{-10pt}
\end{figure}

\noindent\textbf{Data Component}.The conversion rules for data components are as follows: Data components are transformed into the corresponding data types in C++, and data subcomponent instances are converted into variables of the corresponding type. The code generation occurs within the process or thread component files that contain the data components. The conversion mapping shown in Figure \ref{fig14}.

During the component transformation process, there are also conversions involving features, connections, flows, and other AADL elements. For these elements, corresponding conversion rules have been designed, with the implementation approach being similar to the above, and thus will not be elaborated on here.

\subsection{Code Generation Experiment}
\begin{figure}
    \centering
    \includegraphics[width=0.8\textwidth]{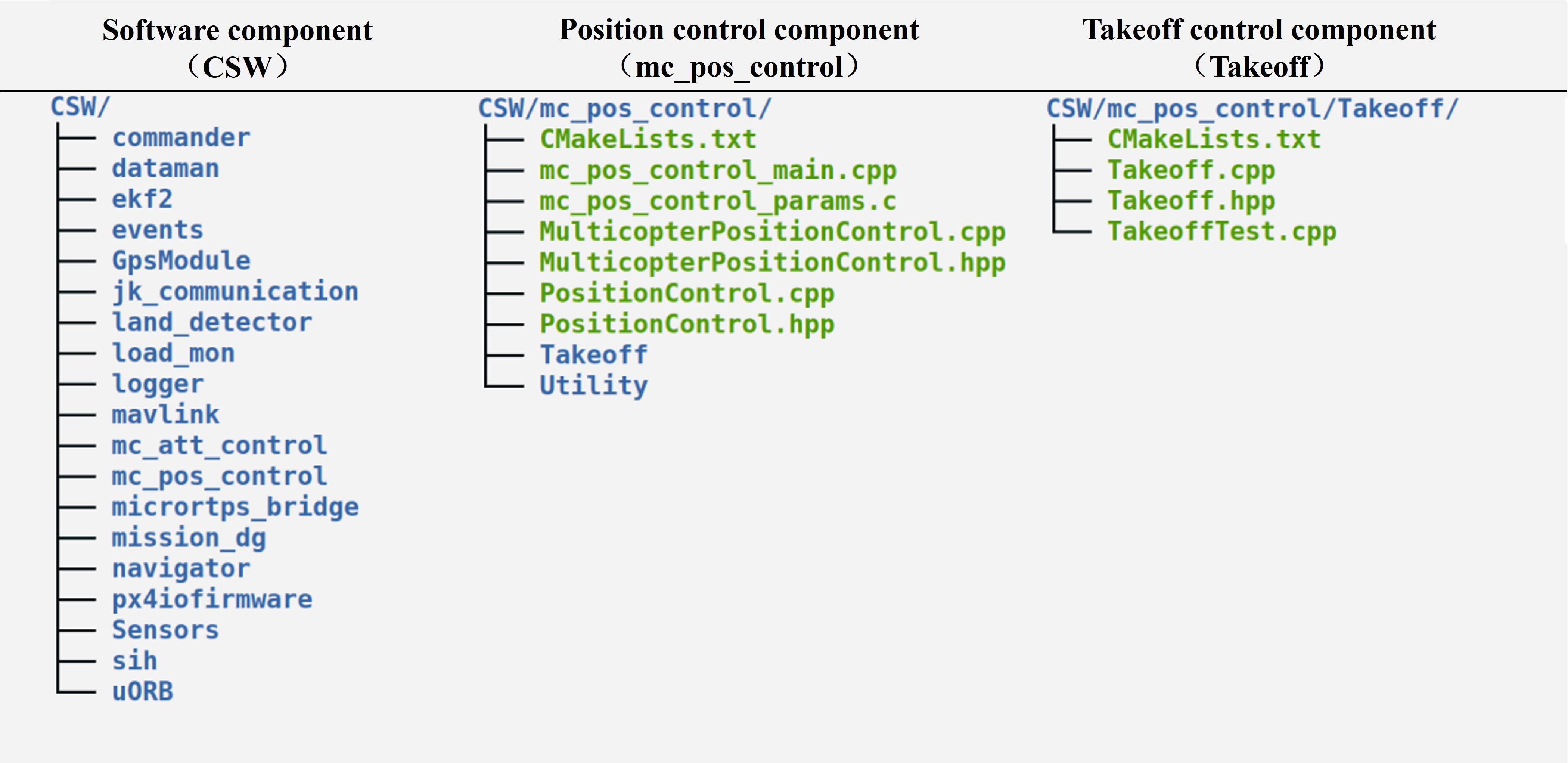}
    \caption{Directory structure of generated flight controller code.}
    \label{fig15}
    \vspace{-10pt}
\end{figure}

This section performs an experiment using the aforementioned code generation method to transform the UAS model. After the transformation, the directory structure of the generated flight controller code is shown in the Figure \ref{fig15}.

The directory structure matches the structure of the software component in the UAS model.

\begin{figure}
    \centering
    \includegraphics[width=0.8\textwidth]{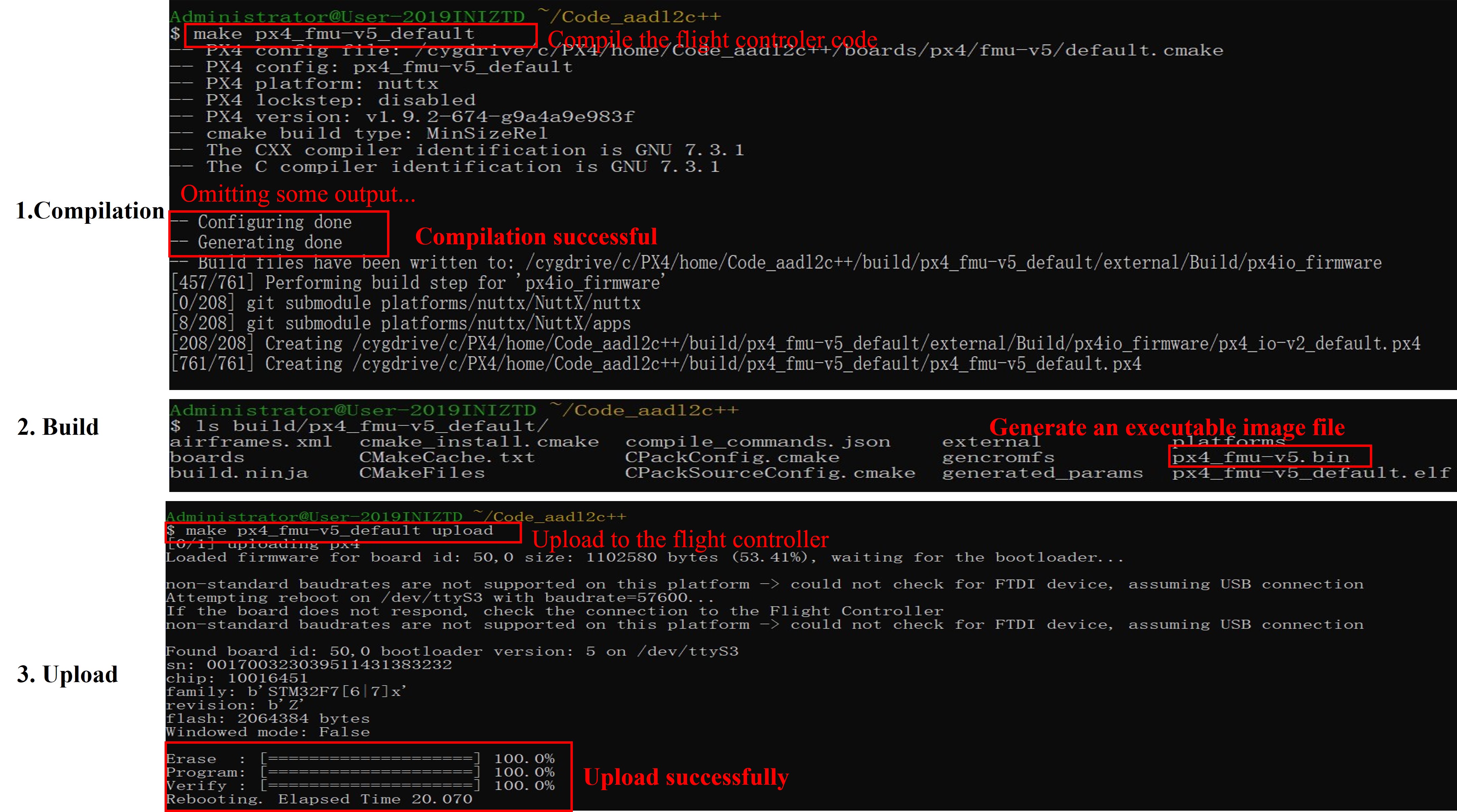}
    \caption{Generated code test.}
    \label{fig16}
    \vspace{-10pt}
\end{figure}

To verify if the code is correct, we successively tested whether the code could be compiled successfully, generate executable image files, and upload to the PX4 flight controller board. The output of the test is shown in the Figure \ref{fig16}.

As illustrated, the code was compiled correctly, generating an executable image file px4\_fmu-v5.bin, and successfully uploaded to the flight controller board. We also used this code to perform drone simulation flights, and during the simulation, the system was able to function normally, indicating that the generated code is viable.

\section{Method Comparison}\label{sec7}
We have compared the method proposed in this paper with existing design-time security assurance technologies for Unmanned Aerial Systems, with the results summarized in Table \ref{table2}.

\begin{table}
    \centering
    \caption{Method Function Comparison.}\
    \includegraphics[width=\textwidth]{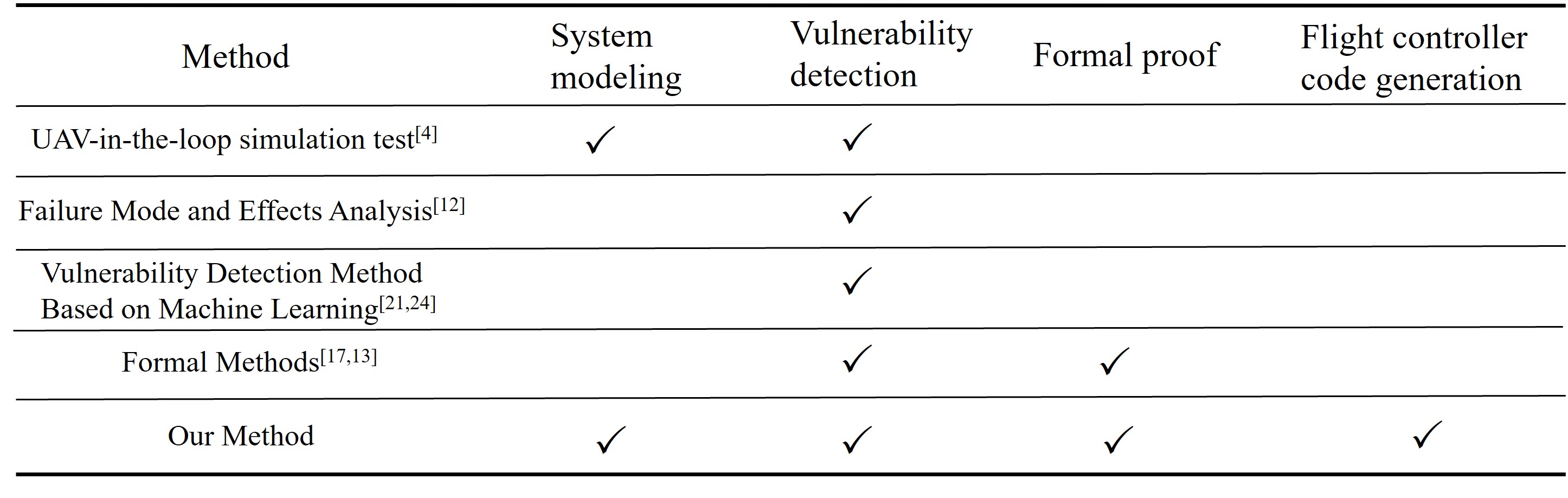}
    \label{table2}
    \vspace{-10pt}
\end{table}

Compared to other methods in the table, the method proposed in this paper is more comprehensive, capable of system modeling, vulnerability detection, formal proof, and flight controller code generation. Unlike in-the-loop simulation testing \cite{dai2021rflysim}, Failure Modes and Effects Analysis \cite{huang2020failure}, and machine learning-based vulnerability detection \cite{sadhu2020board}, our method employs formal techniques for verification, making the reliability proof more complete \cite{clarke1996formal}. Compared to formal methods \cite{luckcuck2023using}, our approach can detect vulnerabilities at a lower cost during the system modeling phase \cite{fisher2017hacms}. Additionally, our method includes the capability to generate flight controller code, transforming verified system models into flight controller code to enhance development efficiency.

\section{Summary and Future Outlook}\label{sec8}

To address the challenges of high time and labor costs associated with the design and verification of highly reliable Unmanned Aerial Systems, this paper investigates an integrated method for modeling, verification, and code generation of UAS. This method combines architecture description languages and formal techniques, offering advantages of lower verification costs and higher development efficiency. However, there are limitations to the current method, as UAS require numerous safety properties, which are difficult to fully describe through manual design during the design phase.

The next phase of research involves leveraging large language model technology to automatically generate safety property specifications. There are existing studies discussing the use of large models for generating formal specifications \cite{kogler2024reliable}, such as using large models to generate safety specifications in the domain of smart contracts \cite{liu2024propertygpt}. We plan to explore the use of large models to assist in generating safety specifications for UAS, thereby enhancing the reliability of UAS.

%
%

%
%
%
\bibliographystyle{splncs04}
\bibliography{mybibliography}

\begin{thebibliography}{10}
\providecommand{\url}[1]{\texttt{#1}}
\providecommand{\urlprefix}{URL }
\providecommand{\doi}[1]{https://doi.org/#1}

\bibitem{balestrieri2021sensors}
Balestrieri, E., Daponte, P., De~Vito, L., Lamonaca, F.: Sensors and measurements for unmanned systems: An overview. Sensors  \textbf{21}(4), ~1518 (2021)

\bibitem{clarke1996formal}
Clarke, E.M., Wing, J.M.: Formal methods: State of the art and future directions. ACM Computing Surveys (CSUR)  \textbf{28}(4),  626--643 (1996)

\bibitem{cofer2017secure}
Cofer, D., Backes, J., Gacek, A., DaCosta, D., Whalen, M., Kuz, I., Klein, G., Heiser, G., Pike, L., Foltzer, A., et~al.: Secure mathematically-assured composition of control models. Air Force Research Laboratory Information Directorate  (2017)

\bibitem{dai2021rflysim}
Dai, X., Ke, C., Quan, Q., Cai, K.Y.: Rflysim: Automatic test platform for uav autopilot systems with fpga-based hardware-in-the-loop simulations. Aerospace Science and Technology  \textbf{114},  106727 (2021)

\bibitem{desai2019soter}
Desai, A., Ghosh, S., Seshia, S.A., Shankar, N., Tiwari, A.: Soter: a runtime assurance framework for programming safe robotics systems. In: 2019 49th Annual IEEE/IFIP International Conference on Dependable Systems and Networks (DSN). pp. 138--150. IEEE (2019)

\bibitem{feiler2004open}
Feiler, P.: Open source aadl tool environment (osate). In: AADL Workshop, paris. pp. 1--40 (2004)

\bibitem{feiler2006architecture}
Feiler, P.H., Gluch, D.P., Hudak, J.: The architecture analysis \& design language (aadl): An introduction  (2006)

\bibitem{ferreira2010unmanned}
Ferreira, S., Medvidovi{\'c}, N., Deonandan, I., Valerdi, R., Hess, J., Mikaelian, T., Shull, G.: Unmanned and autonomous systems of systems test and evaluation: challenges and opportunities  (2010)

\bibitem{fisher2017hacms}
Fisher, K., Launchbury, J., Richards, R.: The hacms program: using formal methods to eliminate exploitable bugs. Philosophical Transactions of the Royal Society A: Mathematical, Physical and Engineering Sciences  \textbf{375}(2104),  20150401 (2017)

\bibitem{gacek2014resolute}
Gacek, A., Backes, J., Cofer, D., Slind, K., Whalen, M.: Resolute: an assurance case language for architecture models. ACM SIGAda Ada Letters  \textbf{34}(3),  19--28 (2014)

\bibitem{gupta2013review}
Gupta, S.G., Ghonge, D.M., Jawandhiya, P.M.: Review of unmanned aircraft system (uas). International Journal of Advanced Research in Computer Engineering \& Technology (IJARCET) Volume  \textbf{2} (2013)

\bibitem{huang2020failure}
Huang, J., You, J.X., Liu, H.C., Song, M.S.: Failure mode and effect analysis improvement: A systematic literature review and future research agenda. Reliability Engineering \& System Safety  \textbf{199},  106885 (2020)

\bibitem{khan2020formal}
Khan, W., Kamran, M., Naqvi, S.R., Khan, F.A., Alghamdi, A.S., Alsolami, E.: Formal verification of hardware components in critical systems. Wireless Communications and Mobile Computing  \textbf{2020},  1--15 (2020)

\bibitem{kogler2024reliable}
Kogler, P., Falkner, A., Sperl, S.: Reliable generation of formal specifications using large language models. In: SE 2024-Companion. pp. 141--153. Gesellschaft f{\"u}r Informatik eV (2024)

\bibitem{lee1999runtime}
Lee, I., Kannan, S., Kim, M., Sokolsky, O., Viswanathan, M.: Runtime assurance based on formal specifications. In: International Conference on Parallel and Distributed Processing Techniques and Applications. pp. 279--287. PDPTA' (1999)

\bibitem{liu2024propertygpt}
Liu, Y., Xue, Y., Wu, D., Sun, Y., Li, Y., Shi, M., Liu, Y.: Propertygpt: Llm-driven formal verification of smart contracts through retrieval-augmented property generation. arXiv preprint arXiv:2405.02580  (2024)

\bibitem{luckcuck2023using}
Luckcuck, M.: Using formal methods for autonomous systems: Five recipes for formal verification. Proceedings of the Institution of Mechanical Engineers, Part O: Journal of Risk and Reliability  \textbf{237}(2),  278--292 (2023)

\bibitem{medvidovic2000classification}
Medvidovic, N., Taylor, R.N.: A classification and comparison framework for software architecture description languages. IEEE Transactions on software engineering  \textbf{26}(1),  70--93 (2000)

\bibitem{meier2015px4}
Meier, L., Honegger, D., Pollefeys, M.: Px4: A node-based multithreaded open source robotics framework for deeply embedded platforms. In: 2015 IEEE international conference on robotics and automation (ICRA). pp. 6235--6240. IEEE (2015)

\bibitem{mohsan2023unmanned}
Mohsan, S.A.H., Othman, N.Q.H., Li, Y., Alsharif, M.H., Khan, M.A.: Unmanned aerial vehicles (uavs): Practical aspects, applications, open challenges, security issues, and future trends. Intelligent Service Robotics  \textbf{16}(1),  109--137 (2023)

\bibitem{sadhu2020board}
Sadhu, V., Zonouz, S., Pompili, D.: On-board deep-learning-based unmanned aerial vehicle fault cause detection and identification. In: 2020 ieee international conference on robotics and automation (icra). pp. 5255--5261. IEEE (2020)

\bibitem{schierman2020runtime}
Schierman, J.D., DeVore, M.D., Richards, N.D., Clark, M.A.: Runtime assurance for autonomous aerospace systems. Journal of Guidance, Control, and Dynamics  \textbf{43}(12),  2205--2217 (2020)

\bibitem{shafiee2021unmanned}
Shafiee, M., Zhou, Z., Mei, L., Dinmohammadi, F., Karama, J., Flynn, D.: Unmanned aerial drones for inspection of offshore wind turbines: A mission-critical failure analysis. Robotics  \textbf{10}(1), ~26 (2021)

\bibitem{taimoor2023novel}
Taimoor, M., Lu, X., Maqsood, H., Sheng, C.: A novel fault diagnosis in sensors of quadrotor unmanned aerial vehicle. Journal of Ambient Intelligence and Humanized Computing  \textbf{14}(10),  14081--14099 (2023)

\bibitem{tan2020unmanned}
Tan, Y., Wang, J., Liu, J., Zhang, Y.: Unmanned systems security: Models, challenges, and future directions. IEEE Network  \textbf{34}(4),  291--297 (2020)

\bibitem{veres2011autonomous}
Veres, S.M., Molnar, L., Lincoln, N.K., Morice, C.P.: Autonomous vehicle control systems—a review of decision making. Proceedings of the Institution of Mechanical Engineers, Part I: Journal of Systems and Control Engineering  \textbf{225}(2),  155--195 (2011)

\bibitem{whalen2012your}
Whalen, M.W., Gacek, A., Cofer, D., Murugesan, A., Heimdahl, M.P., Rayadurgam, S.: Your" what" is my" how": Iteration and hierarchy in system design. IEEE software  \textbf{30}(2),  54--60 (2012)

\bibitem{witayangkurn2012real}
Witayangkurn, A., Nagai, M., Honda, K., Dailey, M., Shibasaki, R.: Real-time monitoring system using unmanned aerial vehicle integrated with sensor observation service. The International Archives of the Photogrammetry, Remote Sensing and Spatial Information Sciences  \textbf{38},  107--112 (2012)

\end{thebibliography}
%





\end{document}